\documentclass[10pt,twocolumn,prb,aps,floats,graphicx,epsf]{revtex4}
\usepackage{amsmath}
\usepackage{graphicx}
\usepackage{color,hyperref}
\hypersetup{colorlinks}
\hypersetup{
	colorlinks=true,
	citecolor=blue
}

\begin{document}

\title{Van der Waals forces from first principles for periodic systems: Application to graphene-water interactions}

\author{Pouya Partovi-Azar}
\affiliation{Department of Chemistry, University of Paderborn, Warburger Str. 100, D-33098 Paderborn, Germany}

\author{Thomas D. K\"{u}hne}
\affiliation{Department of Chemistry and Institute for Lightweight Design with Hybrid Systems, University of Paderborn, Warburger Str. 100, D-33098 Paderborn, Germany}
\email{tdkuehne@mail.upb.de}

\begin{abstract}
We extend the method of Silvestrelli [P. L. Silvestrelli, J. Chem. Phys. {\bf 139}, 054106 (2013)] to approximate long-range van der Waals interactions at the density functional theory level based on maximally localized Wannier functions combined with the quantum harmonic oscillator model, to periodic systems. 
%The van der Waals interactions are taken into account as dynamic correlation effects between the dipole moments attributed to maximally localized Wannier functions, and are estimated using quantum harmonic oscillator model, including contributions beyond pair-wise interactions. 
Applying this scheme to study London dispersion forces between graphene and water layers, we demonstrate that collective many-body effects beyond simple additive pair-wise interactions are essential to accurately describe van der Waals forces. 
%We study dispersion interactions between graphene and a water slab on top, by means of density functional theory calculations. The interaction was also studied in case of bilayer graphene and water slab. We show that the presence of a graphene layer that lies beneath the first one qualitatively changes the van der Waals interactions, and the interactions can not be simply regarded as the sum of interactions between the individual graphene layers and the water slab. 
%Our findings demonstrate the many-body nature of the van der Waals interactions in water-graphene systems, which otherwise is neglected using additive classical treatments. 
\end{abstract}

%\noindent PACS numbers:  81.05.Uw, 71.15.Pd, 82.45.Mp, 73.22.Pr

\maketitle

\section{Introduction}

Graphene, which consists of a single layer of carbon atoms arranged in a $sp^2$ honey-comb structure, \cite{0,1} has been subject of a great number of studies, owing to its unique properties that were experimentally observed and theoretically predicted. These include, but not limited to, electric field \cite{efield} and quantum Hall effects, \cite{2} ultra-high carrier mobility, \cite{3} superior thermal conductivity, \cite{4} high mechanical strength, \cite{5} electron-hole puddles, \cite{7,PPA_PRB} and sensitivity to adsorbents. \cite{schedin} Although, there is a large effort focusing on controlling and adapting graphene for future use in real devices, \cite{8, 9, 10, 10-0, 10-00} some rather fundamental aspects still remain incompletely understood. Among others, is the interaction between graphene sheets and individual water molecules on top of them. \cite{schyman_PCL, Ma_PRB, voloshina_PCCP, huff, rubes, reyes, sudiarta} This interaction can cause a doping effect on graphene and change the density of states near the Fermi level, which therefore is a rather promising effect for potential applications. \cite{wehling} However, weak dispersion interactions play an important role in the binding between carbon atoms and water molecules. Hence, there seems to be a controversy in the reported results, in particular whether or not graphene is transparent with respect to van der Waals (vdW) interactions between the substrate beneath graphene and the water droplet on top of it. \cite{rafiee_Nature_Matter, shih_PRL, shih_nature} Therefore, an accurate treatment of vdW interactions between water and graphene is required. The difficulty of conventional local or semi-local density functional theory (DFT) to quantitatively describe these interactions originates from the long-range nature of dispersion forces. \cite{KlimesMichaelides2012}  Even though a large number of different empirical and \emph{first-principles} techniques have been suggested to calculate vdW interactions within DFT calculations, \cite{Ahlrichs1977, RA1991, Kohn1998, Scheffler1999, Scoles2001, Fuchs2002, WuYang2002, dion, Zimmerli2004, lilienfeld, grimme2004, sebastiani2005, donchev_JCP, WGS2006, Silvestrelli2008, tkatchenko_PRL_2009, Silvestrelli2009, Hesselmann2010, grimme_2010, vydrov, tkatchenko_PRL_2012, Gironcoli} there is still great demand for an accurate and, at the same time, computationally very efficient \emph{first-principles} technique to account for vdW forces in large systems.

Here, we extend the recently introduced method of Silvestrelli \cite{silvestrelli} which combines Quantum Harmonic Oscillator (QHO) model with Maximally Localized Wannier Functions (MLWF), \cite{MarzariVanderbilt1997} to periodic systems. Using this approach, 
%Here, in order to account for dispersion interactions, we use a newly introduced technique, combining the Quantum Harmonic Oscillator (QHO) model with Maximally Localized Wannier Functions (MLWF). \cite{silvestrelli} Although originally introduced for non-periodic systems, here we extend the method to periodic systems by including MLWFs in neighboring cells, in a systematic way. 
%we investigate how the vdW interactions would be different from those obtained from standard DFT and semi-empirical \cite{grimme_2010} calculation. 
we investigate the vdW interactions between an extended layer of water and graphene with respect to the number of layers. The question on the additivity of  vdW interactions between the water and graphene layers is discussed in detail. 

This remaining of the paper is organized as follows. The theoretical method is outlined in Section II, while the computational details are specified in Section III. The eventual results are shown and discussed in Section IV. %before we conclude this article in Section V. 

\section{Method of Calculation}

Considering MLWFs as charge distributions in real-space, it is directly possible to compute the corresponding dipole moments. \cite{Vanderbilt1993} Based on this information, the dynamical electron-correlation effects that arise from many-body instantaneous long-range interactions of the oscillating dipoles can be quantified by the QHO model \cite{donchev_JCP, tkatchenko_PRL_2012, silvestrelli} 
\begin{equation}
H=-\frac{1}{2} \sum_{i=1}^{N} \nabla_{\chi_i}^2 + \frac{1}{2} \sum_{i=1}^{N} \omega_i^2 \chi_i^2
+ \sum_{i > j=1}^{N} \omega_i \omega_j \sqrt{\alpha_i \alpha_j} \boldsymbol{\chi}_i T_{ij} \boldsymbol{\chi}_j,
\end{equation}
where $N$ is the number of MLWFs, while $\boldsymbol{\chi}_i=\sqrt{m_i} \boldsymbol{\zeta}_i$ with $m_i$ being the masses and $\boldsymbol{\zeta}_i$ the displacements of the oscillators from equilibrium. The frequency and polarizability of the oscillators are denoted as $\omega_i$ and $\alpha_i$, while $T_{ij}$ is the dipole-dipole interaction tensor. Knowing the centers and spreads $S_i$ of the MLWFs, the respective polarizabilities $\alpha_i \sim \gamma S_i^3$ and characteristic frequencies $\omega_i \sim \sqrt{Z_i/\alpha_i}$ can be calculated, where $\gamma$ is a proportionality constant and $Z_i$ the atomic number. \cite{silvestrelli, ambrosetti} Moreover, $T_{ij}$ is modified to allow for orbital overlap at short distances, by introducing a short-range damping function for the bare Coulomb potential. \cite{silvestrelli} The energies of all $N$ 3-dimensional QHOs can be found by diagonalizing a $3N \times 3N$ matrix $\mathbf{C}$ that is defined as
\begin{subequations}
\begin{eqnarray}\label{cmatrix}
C_{ii}&=&\omega_i^2 \\ %\boldsymbol{I}; 
C_{i \neq j} &=& \omega_i \omega_j \sqrt{\alpha_i \alpha_j} T_{ij},
\end{eqnarray}
\end{subequations}
which contains $N^2$ $3 \times 3$ matrix-blocks corresponding to the individual MLWFs. %$\boldsymbol{I}$ is the identity matrix. 
The vdW correction can then be obtained by
\begin{equation}\label{ec}
E_{vdW}=\frac{1}{2} \sum_{p=1}^{3N} \sqrt{\lambda_p} - \frac{3}{2} \sum_{i=1}^{N} \omega_i,
\end{equation}
where $\lambda_i$ are eigenvalues of the correlated system, while $\omega_i$ are the aforementioned characteristic frequencies of the dipole moments attributed to the MLWFs. 

Due to the fact that $T_{ij}$ decays relatively quickly with respect to the distance between the MLWFs, instead of a genuine Ewald sum, the interactions between the MLWFs in the unit cell and those in its periodic images are taken into account by considering a finite buffer region around the unit cell, where the MLWFs of the original unit cell are replicated in all spatial directions. To that extend we decompose the vdW interaction energy for the extended system $E_{vdW}^{ext}$ into a sum of vdW interaction energies of the MLWFs in the unit cell $E_{vdW}^{UC-UC}$, the interaction energy between the MLWFs in the unit cell and those in the buffer zone $E_{vdW}^{UC-b}$, and interaction energy between the MLWFs in the buffer region only, which is denoted as $E_{vdW}^{b-b}$. The vdW interaction energy of the extended system then reads as
\begin{equation}
E_{vdW}^{ext}=E_{vdW}^{UC-UC}+E_{vdW}^{UC-b}+E_{vdW}^{b-b},
\end{equation}
from which the last term has to be subtracted to yield the desired vdW interaction energy of the original system, i.e.
\begin{equation}
E_{vdW}^{tot}=E_{vdW}^{ext}-E_{vdW}^{b-b}. \label{TotEner}
\end{equation}
As a consequence, the total energy including the vdW correction is $E^{tot} = E_{DFT}^{tot} + E_{vdW}^{tot}$, $E_{DFT}^{tot}$ is the total DFT energy. 

\section{Computational Details}

\begin{figure}
\includegraphics[height=5.0cm]{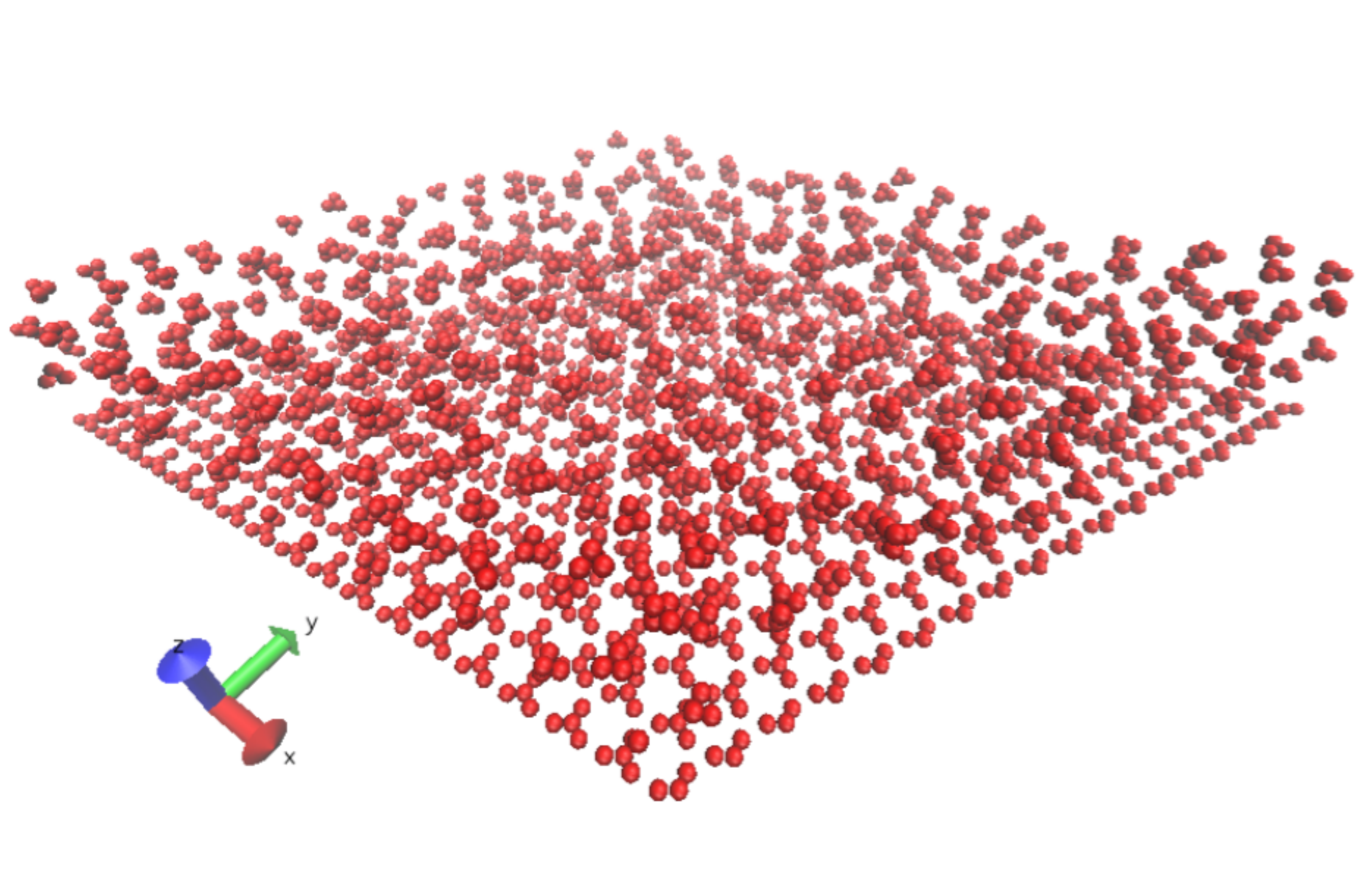}
\caption{The MLWFs centers of the extended system (unit cell plus additional buffer region) of the SLG-water system at 3~$\text{\AA}$ vertical separation.}\label{fig1}
\end{figure}

In the following we have considered two systems, a single layer of graphene (SLG), as well as bilayer graphene (BLG), both with a 100 molecule water slab on top. The graphene, which consisted of 128 carbon atoms per layer, was placed in a periodic orthorhombic simulation box parallel to the $xy$-plane with a large 35~$\text{\AA}$ vacuum portion along the perpendicular $z$-direction. The MLWF centers of the eventual system are shown in Fig.~\ref{fig1}

In order to obtain the atomic configuration of the water molecules, the {\it ab-initio} molecular dynamics (AIMD) simulation were performed in the canonical ensemble at 300~K using the second-generation Car-Parrinello method of K\"uhne et al. \cite{TDK, CP2G} as implemented in the Gaussian and plane wave \cite{Lippert} DFT code CP2K/\textsc{Quickstep}. \cite{cp2k} In AIMD simulation carbon atoms of graphene where fixed. The interatomic interactions were described by DFT \cite{DFTrev} employing the Perdew-Burke-Ernzerhof (PBE) exchange-correlation functional \cite{PBE}, Goedecker-Teter-Hutter pseudopotentials \cite{GTH, Krack2005} and a double-$\zeta$ Gaussian basis set with one addition al set of polarization functions. \cite{VandeVondeleGaussian} The spread of the Wannier orbitals was minimized using the scheme of Berghold et al. \cite{berghold}

\begin{figure}
\centerline{\includegraphics[width=8.0cm]{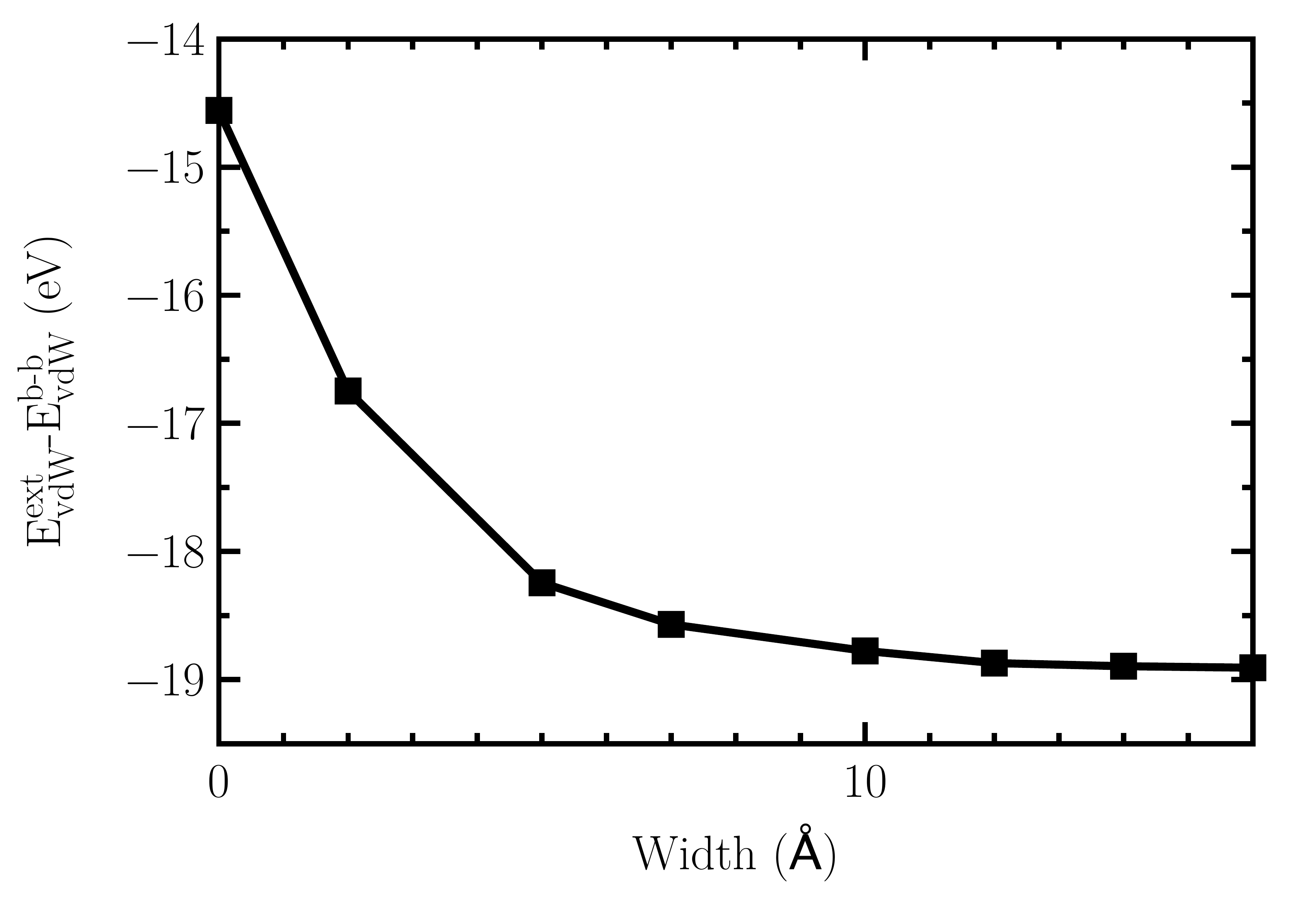}}
\caption{The vdW energy as a function of buffer region width.}\label{fig2}
\end{figure}
To estimate the size of the buffer zone in Fig.~\ref{fig2} the vdW energy as defined in Eq.~\ref{TotEner} is shown as a function of additional buffer width, $d$. We found that d=12~$\text{\AA}$ is sufficient to adequately converge the vdW and to capture most of the relevant dispersion interactions, while at the same time keeping the size of the $\mathbf{C}$ matrix manageable.

\section{Results and Discussion}

\begin{figure}
\centerline{
\includegraphics[height=6.0cm]{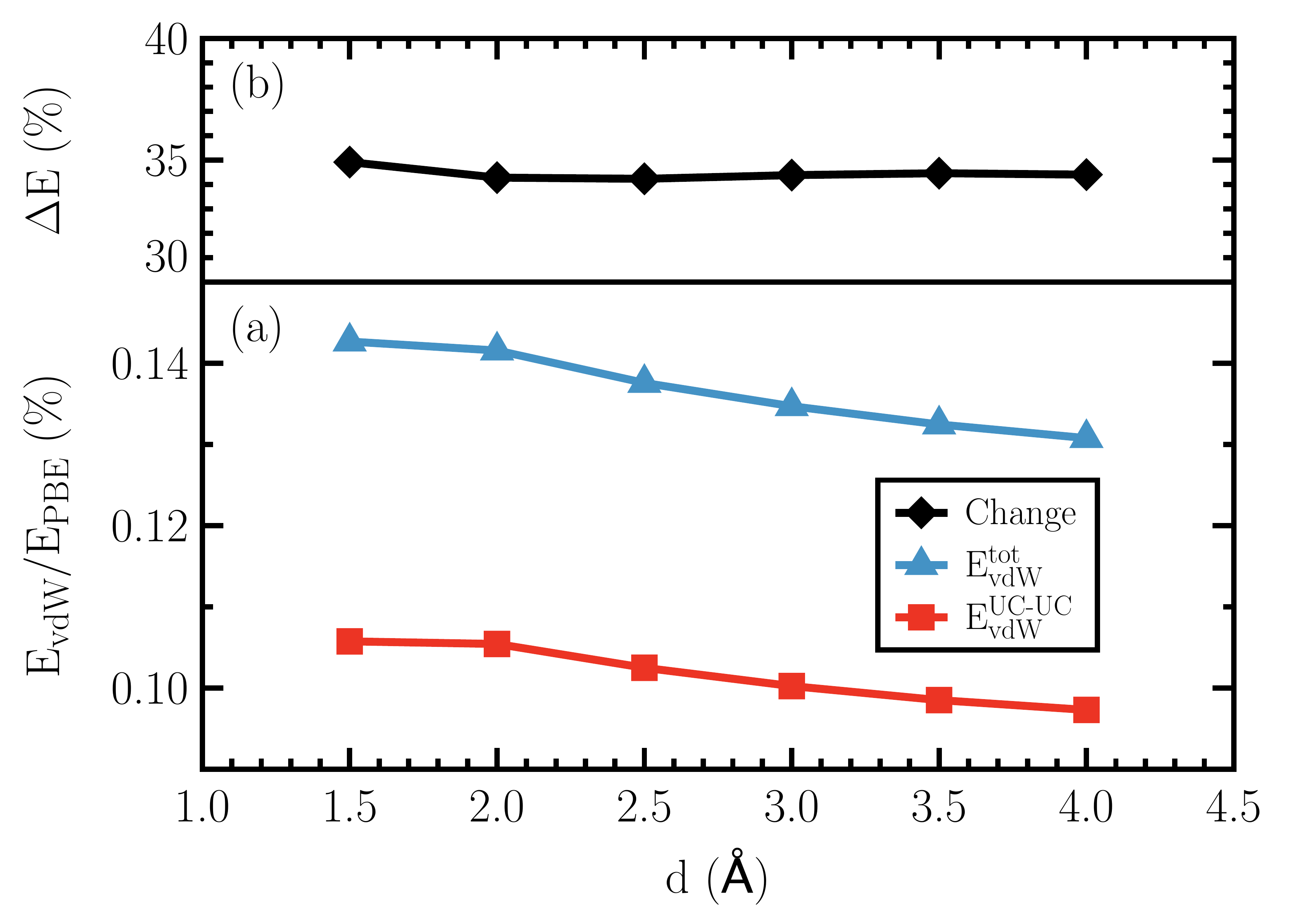}}
\caption{(a) The vdW interaction energy as a function of distance between the water slab and the SLG with (blue triangles) and without (red squares) the MLWFs of the buffer zone.
(b) Relative difference between $E_{vdW}^{UC-UC}$ and $E_{vdW}^{tot}$.}\label{fig3}
\end{figure}

To demonstrate the impact of periodic boundary conditions (PBC), Fig~\ref{fig3}(a) displays the vdW energy as a function of the distance between the water slab and SLG, which is defined as the vertical gap between the nearest H atom of the water slab and the $xy$-plane. 
As can be seen in Fig.~\ref{fig3}(b), the vdW energy with and without our correction for PBC differ by $\sim$34\%. 
\begin{figure}
\centerline{
\includegraphics[height=6.0cm]{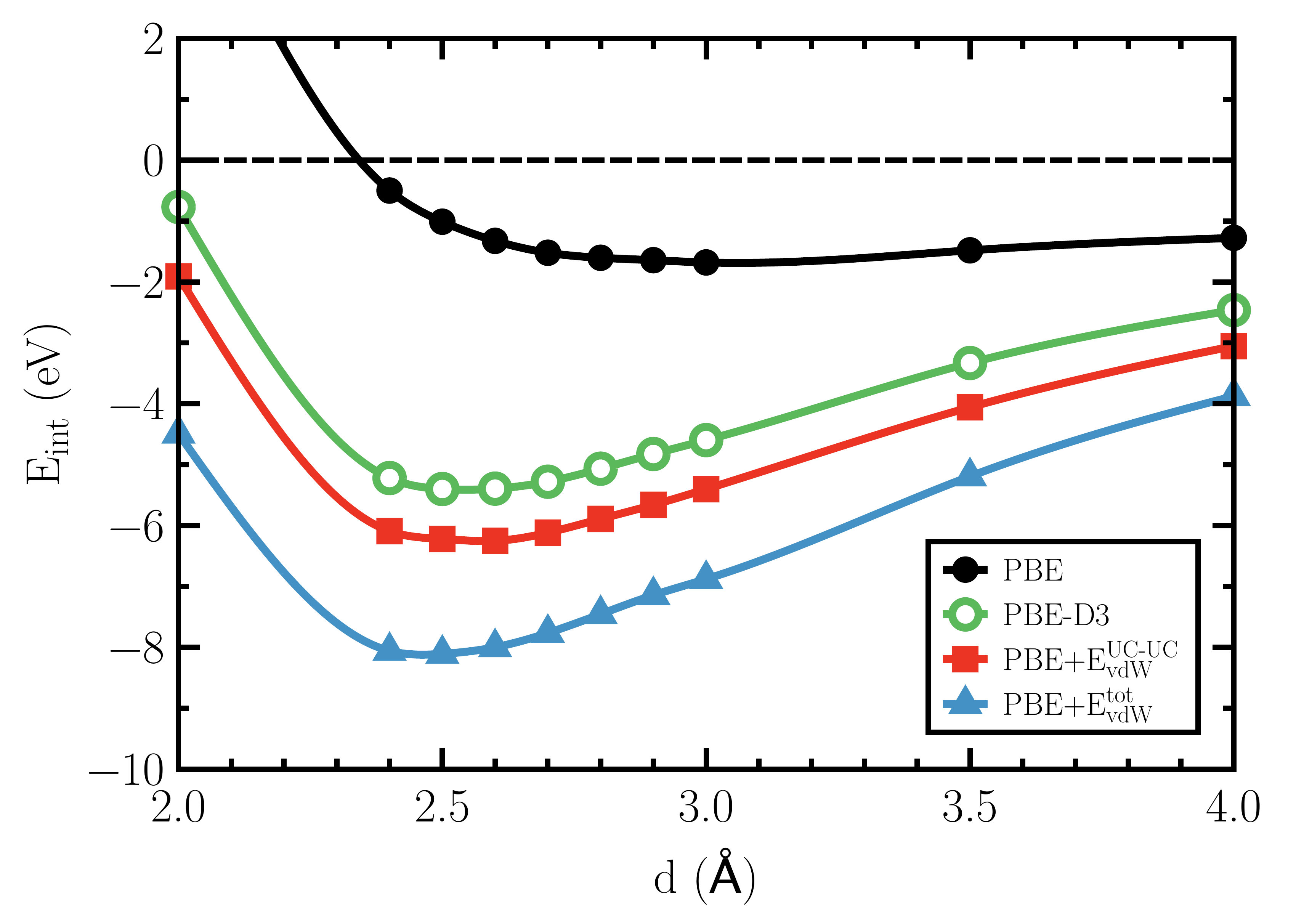}}
\caption{Interaction energy between the water slab and the SLG as a function of their distance, using PBE without vdW correction (black circles), PBE-D3 (green circles), PBE+$E_{vdW}^{UC-UC}$ (red squares), and PBE+$E_{vdW}^{tot}$ (blue triangles).}\label{fig4}
\end{figure}
In Fig.~\ref{fig4} the interaction energy $E_{int}$ as obtained from bare PBE, PBE-D3, PBE+$E_{vdW}^{UC-UC}$ and PBE+$E_{vdW}^{tot}$ are shown as a function of the vertical distance between the water slab and SLG. It is apparent that without any vdW correction PBE hardly binds, with the associated equilibrium distance being about 3.0~$\text{\AA}$. The corresponding equilibrium distance for both, PBE-D3 and PBE+$E_{vdW}^{tot}$ are 2.5~$\text{\AA}$ and 2.6~$\text{\AA}$ for PBE+$E_{vdW}^{UC-UC}$, which is in good agreement with previous results obtained by others using a large variety of different methods such as polarizable and non-polarizable force fields, DFT calculations including dispersion corrections, calculations based on the random phase approximation, but also with local MP2 and CCSD(T) calculations of a single water molecule on a hydrogen-terminated graphene flake. \cite{schyman_PCL, Ma_PRB, voloshina_PCCP, huff, rubes, reyes, sudiarta} However, at variance to the latter, here a disordered water slab is considered, which comprises a large number of different water orientations towards the graphene layer. Nevertheless, for most of the water molecules, one O-H bond is preferably pointing towards the hydrophobic surface, as has been observed in previous AIMD simulations. \cite{galli, kuehne_JPCL} The eventual interaction energy averaged over all water molecules is approximately $82 \ \text{meV}$ per water molecule.

\begin{figure}
\centerline{
\includegraphics[height=6.0cm]{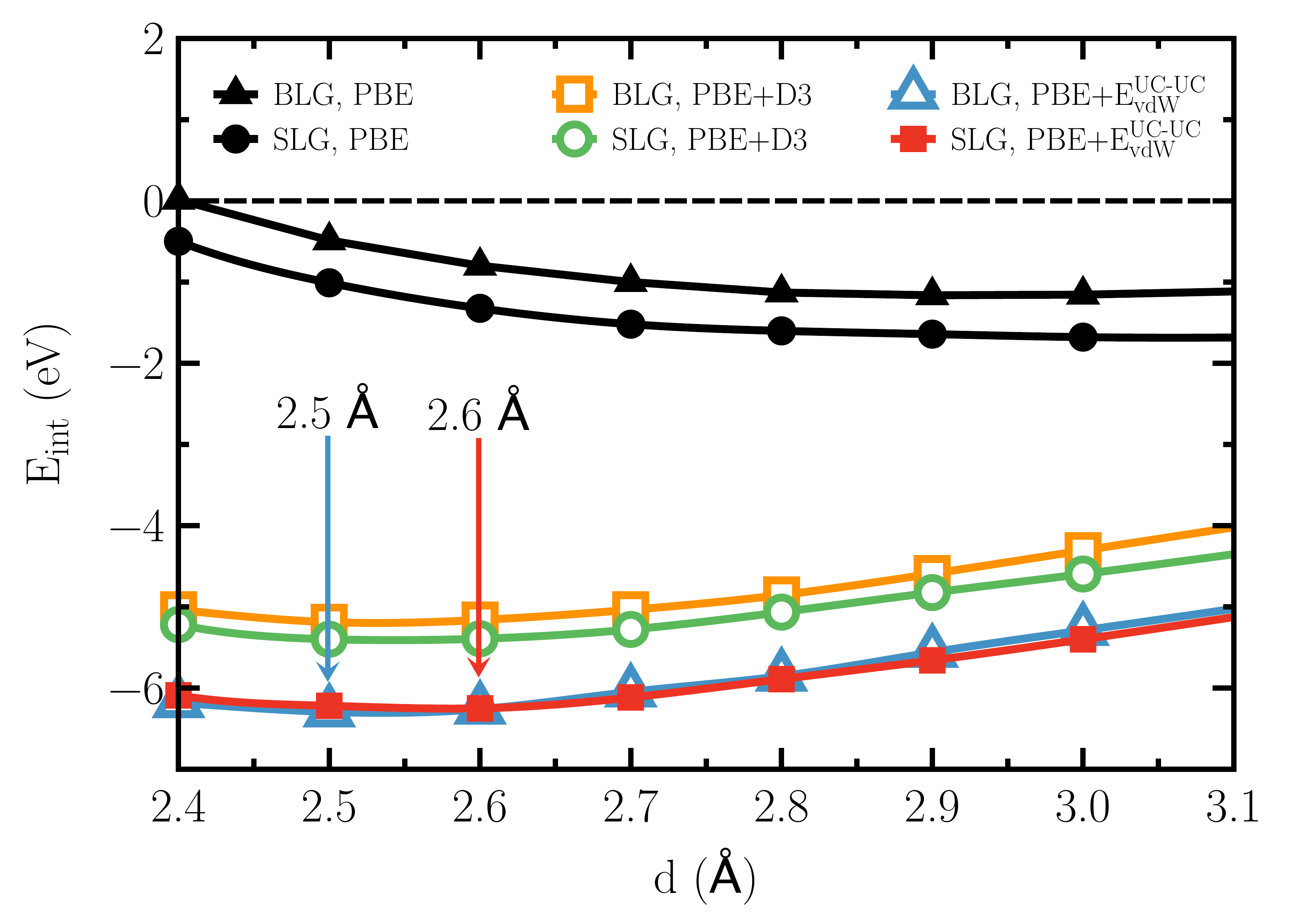}}
\caption{Interaction energy between the water slab and SLG/BLG as a function of distance between the upper graphene and the water layer. The PBE results for the SLG (BLG) are shown in solid black circles (triangles), while the PBE-D3 calculations are denoted by green (orange) spheres (squares). The corresponding biding curves in case of the present $E_{vdW}^{UC-UC}$ correction for SLG and BLG are depicted as red squares and blue triangles, respectively.}\label{fig5}
\end{figure}

The results for the interaction energy $E_{int}$ between the water slab and BLG with AB stacking at the experimentally observed separation of 3.34~$\text{\AA}$, \cite {gr_spa_1,gr_spa_2,gr_spa_3} are shown in Fig.~\ref{fig5} and compared with the values for SLG. 
Due to the fact that considering BLG including the buffer region would results in a rather large $\mathbf{C}$ matrix that needs to be diagonalized, we have confined ourselves to vdW correction due to MLWFs in the unit cell. As before, employing the PBE functional without any vdW correction, water and BLG barely binds. More interesting, we find that in comparison to the SLG system, $E_{int}$ is reduced by about $\sim$30\%. However, in the case of PBE-D3 calculation, the reduction of $E_{int}$ due to the second graphene layer is just $\sim$5 \%, % when the PBE results are corrected with semi-empirical D3 scheme, while again showing a less binding in the case of the BLG. 
while for $E_{vdW}^{UC-UC}$ the binding energies between SLG and BLG systems are essentially identical. Nevertheless, in the latter case, the equilibrium distance changes from 2.6~$\text{\AA}$ for SLG to 2.5~$\text{\AA}$ for BLG. The PBE-D3 calculations predict an equilibrium distance of 2.5~$\text{\AA}$ for both of the considered systems, while for the bare PBE functional, the equilibrium distance reduces from 3.0~$\text{\AA}$ to 2.9~$\text{\AA}$ for SLG and BLG, respectively. In any case, for all the computational methods we have considered here, the binding energy between the water slab and SLG is at least at large as for BLG. This is to say, that the vdW interactions are non-additive and in fact are screened by the additional graphene layer, which immediately suggest that the electronic structure of an individual sheet is changed dramatically when interacting with other layers. 
To that extend we have calculated the $z$-component of the molecular dipole moment of both graphene systems without water using the Berry phase operator for periodic systems and the centers of the MLWFs. \cite{Vanderbilt1993, MarzariVanderbilt1997} In the case of the SLG, the $z$-component of the dipole is unsurprisingly zero, while in the case of BLG, each layer exhibits a dipole moment of $\sim$3.65~D, though in opposite directions. %The complex cooperatively effect in the hydrogen bond network of the water slab comes on top, presumably affecting mostly the nearest graphene layer.
We conclude by noting that the latter is a manifestation that the whole is more than the sum of its constituents, which highlights the importance to explicitly consider the electronic structure of the full interacting system to embrace the subtle many-body effects of the vdW interaction. \cite{Vignale2014}

%\section{Conclusions}

%In summary, we have performed DFT calculations while taking the long-range dispersion interactions into account using maximally localized Wannier functions and the quantum harmonic oscillator model, aiming to study the van der Waals interactions between graphene layers and water slab with many-body interactions included. The method has been extended to be also applicable for periodic systems. It provides substantial improvements to the standard PBE calculation, giving, for example, an equilibrium distance of $\sim$2.5~$\text{\AA}$ between the single layer graphene and the water slab and a binding energy of $\sim$82 meV per water molecule, instead of $\sim$3.0~$\text{\AA}$ and $\sim$17 meV which were obtained using pure PBE. We also considered a water slab on top of a bilayer graphene with AB stacking. %We have shown that the cooperativity in essentially long-range dispersion interactions coming from different parts of the system, makes it hard to consider the contribution of each individual part separately. This, of course, arises from the fact that the charge distribution in different parts of the system is profoundly affected by the other parts, specially by the polar ones. And this, in turn, modifies the local dipole moments of the systems, and of course any interactions depending on dipole-dipole coupling. As such, the whole system need to be considered as one entity, and therefore, classical approaches to simulate van der Waals interactions most probably fail to provide a realistic picture.

%\begin{acknowledgments}

%\end{acknowledgments}

\end{document}